\documentclass[pre,twocolumn,showpacs,floatfix]{revtex4}
\usepackage{graphicx}

\def\be{\begin{equation}}
\def\ee{\end{equation}}

\begin{document}
\title{Conditional Lagrangian acceleration statistics in turbulent
flows with Gaussian distributed velocities}
 \author{A.K. Aringazin}
 \email{aringazin@mail.kz}
 \altaffiliation[Also at ]
  {Department of Mechanics and Mathematics, Kazakhstan Division, Moscow State
University, Moscow 119899, Russia.}
 \affiliation{Department of Theoretical Physics, Institute for
Basic Research, Eurasian National University, Astana 473021,
Kazakhstan}

\date{April 17, 2003}

\begin{abstract}
The random intensity of noise approach to one-dimensional
Laval-Dubrulle-Nazarenko type model having deductive support from
the three-dimensional Navier-Stokes equation is used to describe
Lagrangian acceleration statistics of a fluid particle in
developed turbulent flows. Intensity of additive noise and cross
correlation between multiplicative and additive noises entering a
nonlinear Langevin equation are assumed to depend on random
velocity fluctuations in an exponential way. We use exact analytic
result for the acceleration probability density function obtained
as a stationary solution of the associated Fokker-Planck equation.
We give a complete quantitative description of the available
experimental data on conditional and unconditional acceleration
statistics within the framework of a single model with a single
set of fit parameters. The acceleration distribution and variance
conditioned on Lagrangian velocity fluctuations, and  the marginal
distribution calculated by using independent Gaussian velocity
statistics are found to be in a good agreement with the recent
high-Reynolds-number Lagrangian experimental data. The fitted
conditional mean acceleration is very small, that is in agreement
with DNS, and increases for higher velocities but it departs from
the experimental data, which exhibit anisotropy of the studied
flow.
\end{abstract}


\maketitle

\section{Introduction}

\noindent {Data analysis and modeling of statistical properties of
a Lagrangian particle advected by a fully developed turbulent flow
is of much practical interest and complement traditional studies
made in Eulerian framework. Strong and nonlocal character of
Lagrangian particle coupling due to pressure effects makes the
main obstacle to derive turbulence statistics from the
Navier-Stokes equation. Recent breakthrough Lagrangian
experiments~\cite{Bodenschatz,Bodenschatz2,Mordant0303003} have
motivated growing interest to a single-particle statistics.} Some
phenomenological approaches~\cite{Aringazin0204359,Beck3}
{inspired by the non-extensive statistics formalism} were
used~\cite{Beck4-Beck} to describe Lagrangian acceleration of a
fluid particle in a stationary developed turbulent flow within the
framework of Langevin type equation; see also
Refs.~\cite{Sawford-Wilk-Beck2,ReynoldsPF2003,ReynoldsPRL2003}.

Some toy models of developed turbulence suffer from the lack of
justification of a fit from turbulence
dynamics~\cite{Kraichnan0305040}, and the connection between
specific non-thermodynamical processes and non-extensive
mechanisms was argued to be generally not well
defined~\cite{Zanette2004}. Recent one-dimensional (1D) stochastic
particle models and their
refinements~\cite{Aringazin0212462,Aringazin0301040,Aringazin0301245}
were reviewed in Ref.~\cite{Aringazin0305186}, in which importance
of Navier-Stokes equation based approaches is emphasized.

{Fluid particle dynamics in a developed turbulent flow is
described in terms of a generalized Brownian motion with the
Lagrangian acceleration of individual particle viewed as a
dynamical variable. In the data processing, the acceleration is
associated with the Lagrangian velocity increment in time for
sufficiently small timescales, in a far dissipative subrange where
turbulent fluctuations are smoothed. Such models are generally
based upon a hierarchy of characteristic timescales in the system
and naturally employ one-point statistical description using
Langevin type equation, or the associated Fokker-Planck equation
for one-point probability density function of the variable. Noises
entering the Langevin type equation are treated along a fluid
particle trajectory, and the Fokker-Planck approximation makes
connection between the dynamics and statistical approach.}

Recently we have
shown~\cite{Aringazin0305186,Aringazin0305459,Aringazin0306022,Aringazin0311098}
that the 1D Laval-Dubrulle-Nazarenko (LDN) toy
model~\cite{Laval0101036,LavalPF2003} of the acceleration
evolution with the model turbulent viscosity $\nu_{\mathrm t}$ and
coupled delta-correlated Gaussian multiplicative and additive
noises is in a good agreement with the high-precision Lagrangian
experimental data on acceleration
statistics~\cite{Bodenschatz,Bodenschatz2,Mordant0303003}; Taylor
microscale Reynolds number $R_\lambda=690$, the measured
normalized acceleration range is $-60\leq a/\langle
a^2\rangle^{1/2}\leq 60$, and the resolution is 1/65 of the
Kolmogorov length scale of the flow. The longstanding
Heisenberg-Yaglom scaling of a component of Lagrangian
acceleration, $\langle a^2\rangle = a_0{\bar
u}^{9/2}\nu^{-1/2}L^{-3/2}$, was confirmed
experimentally~\cite{Bodenschatz} to a very high accuracy, for
about seven orders of magnitude in the acceleration variance, or
two orders of the root mean square velocity $\bar u$, at
$R_\lambda>500$; {$a_0$ is the Kolmogorov constant, $\nu$ is the
kinematic viscosity, and $L$ is the Lagrangian integral length
scale}. Long-time correlations and the occurrence of very large
fluctuations at small scales dominate the motion of a fluid
particle, and this leads to a new dynamical picture of
turbulence~\cite{Mordant0206013,Chevillard0310105}.

The original 3D and 1D LDN models were formulated both in the
Lagrangian and Eulerian frameworks for small-scale velocity
increments in time and space respectively. They are based on the
Gabor transformation (Fourier transform in windows) and a
stochastic kind of Batchelor-Proudman rapid distortion theory
(RDT) approach to the incompressible 3D Navier-Stokes
equation~\cite{Laval0101036}, and thus have a deductive support
from turbulence dynamics. A study based on direct numerical
simulations (DNS) of the 3D LDN model in the regime of decaying
turbulence has been made.

The random intensity of noise (RIN)
approach~\cite{Aringazin0305186,Aringazin0305459} provides an
extension of the 1D LDN model viewed in the limit of small
timescale $\tau$ for which Lagrangian velocity increments are
proportional to $\tau$: $u(t+\tau)-u(t) = \tau a(t)$.

The main idea of the RIN approach is simply to account for the
recently established two well separated Lagrangian autocorrelation
time scales for the velocity increments~\cite{Mordant0206013} and
assume that certain model parameters, such as intensity of noise,
fluctuate at {\em large} timescale.

An analysis of such a simple 1D model can shed some light to
properties of the 3D LDN model of Lagrangian dynamics. Recent
development of the 3D LDN model can be found in
Ref.~\cite{Dubrulle0304035}, in which some new methods of
turbulent dynamo problem have been exploited.

The experimental data on the {axial} component of Lagrangian
acceleration $a$ of polystyrene tracer particle in the
$R_\lambda=690$ water flow generated between counter-rotating
disks have been fitted by the stretched
exponential~\cite{Bodenschatz,Bodenschatz2,Mordant0303003},
\be\label{Pexper}
P(a) = C \exp\left[-\frac{a^2}{(1
+\left|{b_1a}/{b_2}\right|^{b_3})b_2^2}\right].
\ee
Here, $b_1=0.513\pm 0.003$, $b_2= 0.563\pm 0.02$, and $b_3=
1.600\pm 0.003$ are fit parameters, and $C=0.733$ is a
normalization constant. {The studied flow is highly anisotropic at
large scales and this appears to affect small scales, resulting in
a small skewness of the acceleration distribution and observable
distinction in the distributions of different components of the
velocity.} At large acceleration values, tails of the distribution
(\ref{Pexper}) decay asymptotically as $\exp[-|a|^{0.4}]$ that
implies a convergence of the fourth-order moment $\langle
a^4\rangle =\int_{-\infty}^{\infty} a^4P(a)da$, as confirmed by
the experiment {with measured normalized acceleration values up to
$|a|=60$}. The flatness factor of the distribution (\ref{Pexper})
which characterizes widening of its tails (when compared to a
Gaussian) is $F\equiv {\langle a^4\rangle}/{\langle a^2\rangle^2}
= 55.1$, that is in agreement the experimental value $F=55 \pm 8$.
The Kolmogorov time of the flow is $\tau_\eta=0.93$ ms. Low-pass
filtering with the 0.23$\tau_\eta$ width of the collected
$1.7\times 10^8$ data points was used, and the response time of
the optically tracked 46~$\mu$m tracer particle is
0.12$\tau_\eta$.

Recently, Chevillard {et al.}~\cite{Chevillard0310105} have
constructed an appropriately recasted multifractal approach to
describe statistics of Lagrangian velocity increments in a wide
range of timescales, from the integral to dissipative one. The
resulting theoretical distribution reproduces continuous widening
of the velocity increment probability density function (PDF) with
the decrease of timescale, from a Gaussian-shaped to the stretched
exponential as observed in Lagrangian experiments carried out at
Cornell~\cite{Bodenschatz,Bodenschatz2,Mordant0303003} and
ENS-Lyon~\cite{Mordant0103084,Mordant0206013}, and DNS of the 3D
Navier-Stokes equation. Two global parameters (Reynolds number and
Lagrangian integral timescale) and two local parameters
(intermittency parameter and smoothing parameter) with a parabolic
singularity spectrum were used to cover the data in the entire
range of timescales. At dissipative timescale the obtained PDF
fits the experimental data on Lagrangian acceleration to a good
accuracy. The cumulant analysis made in this approach provides an
understanding of the observed departures from the scaling when
going from the integral to dissipative timescale. The used
parabolic singularity spectrum $D(h)$ is a hallmark of the
log-normal (Kolmogorov 1962) statistics and reproduces well the
left-hand-side (corresponding to intense velocity increments) of
the observed curve which is centered at 0.58 ($>1/2$) but
increasingly deviates at the right-hand-side of it (corresponding
to weak velocity increments). Another widely used statistics, the
log-Poisson one, was shown to depart from Lagrangian observations
in the same manner.
The conditional acceleration statistics was not considered in this
work.

In a recent paper A.~Reynolds~\cite{ReynoldsNEXT2003} developed a
self-consistent second-order stochastic model with additive noise
which accounts for dependence of the Lagrangian acceleration
covariance matrix on Lagrangian velocities $u$. The observed
dependence of the conditional acceleration variance $\langle
a^2|u\rangle$ on $u$~\cite{Mordant0303003} was partially
understood in terms of Lagrangian accelerations induced by vortex
tubes within which the vorticity is constant and outside which the
vorticity vanishes. Scaling relations were invoked to derive a
third-order polynomial structure of the isotropic covariance
matrix as a function of squared velocity
$u^2$~\cite{SawfordPF2003}. The inclusion of such conditional
acceleration covariances in the model resulted in reduction of the
predicted occurrence of small accelerations that meets
experimental and DNS data for unconditional distributions. The
cores of the resulting conditional acceleration distributions were
found to broaden with increasing $u$, in a qualitative agreement
with the experiment.

Sawford {et al.}~\cite{SawfordPF2003} have studied acceleration
statistics from laboratory measurements and direct numerical
simulations in 3D turbulence at $R_\lambda$ ranging from 38 to
1000. For large $|u|$, the conditional acceleration covariance
behaves like $u^6$. This is qualitatively consistent with the
stretched exponential tails of the unconditional acceleration PDF.
The conditional mean rate of change of the acceleration derived
from the data has been shown consistent with the drift term in
second-order Lagrangian stochastic models of turbulent transport.
The correlation between the square of the acceleration and the
square of the velocity has been shown small but not negligible.

{In very recent papers
Biferale~et~al.~\cite{Biferale0402032,Biferale0403020} have
presented interesting results of DNS of Lagrangian transport in
homogeneous and isotropic turbulence with $R_\lambda$ up to 280, a
very accurate resolution of dissipative scales, and an integration
time of about Lagrangian timescale. They have shown how the
multifractal formalism offers an alternative approach which is
rooted in the phenomenology of turbulence. The Lagrangian
statistics was derived from the Eulerian statistics without
introducing {ad hoc} hypotheses. Although the formalism is not
capable to account for small acceleration values (typical
situation for the multifractal approach), the obtained
acceleration PDF captures the DNS data well in the tails, with
normalized acceleration values ranging from about $|a|/\langle
a^2\rangle^{1/2}=1$ up to $|a|/\langle a^2\rangle^{1/2}=80$. Alas,
one can observe an overestimation in this range which can be
clearly seen from the predicted contribution to the fourth-order
moment, $a^4P(a)$, as compared to the DNS data. High degree of
isotropy of the simulated stationary flow suggests equivalence of
Cartesian components of acceleration aligned to fixed directions,
and the resulting DNS distribution obtained by averaging over the
components has been found with no observable asymmetry with
respect to $a\to -a$. The multifractal approach has been also
used~\cite{Biferale0403020} to obtain acceleration moments
conditional on the velocity. Particularly, the multifractal
prediction $\langle a^2|u\rangle \sim u^{4.57}$ agrees well with
the DNS data for large velocity magnitudes. The predicted exponent
4.57 differs from the value 6 predicted recently by Sawford
et~al.~\cite{SawfordPF2003} and is very close to the
Heisenberg-Yaglom scaling exponent value 9/2. This indicates that
the averaging of the above conditional acceleration variance
$\langle a^2|u\rangle$ over Gaussian distributed velocity $u$ is
consistent with the Heisenberg-Yaglom scaling law (see remark and
Eq.~(68) of Ref.~\cite{Aringazin0305186}).}

In the present paper, we focus on 1D LDN type dynamical modeling
of the Lagrangian acceleration {\em conditional on velocity
fluctuations} presented recently by Mordant, Crawford, and
Bodenschatz in the experimental work~\cite{Mordant0303003}. In
contrast to our previous studies of the conditional acceleration
statistics~\cite{Aringazin0305186,Aringazin0305459,Aringazin0306022,Aringazin0311098},
here we give a self-consistent treatment of the model by explicit
accounting for a Gaussian distribution of Lagrangian velocity
fluctuations that is observed experimentally. We give a complete
quantitative description of the available experimental data on
{{\em both} the} conditional and unconditional acceleration
statistics within the framework of a single model with a single
set of fit parameters. {Importance of the present approach is that
the Lagrangian single-particle modeling is dynamical and has a
deductive support from the Navier-Stokes equation, with few
assumptions justified by the turbulence phenomenology being used.
This approach adds a new look to homogeneous isotropic turbulence
modeling which is alternative to those given by the recent
multifractal and ad hoc Langevin stochastic approaches.}

{It should be emphasized that the Lagrangian velocity is known to
follow Gaussian distribution to a very good accuracy while the
Lagrangian acceleration follows highly non-Gaussian distribution
which is related to extremely intermittent character of the
acceleration, with pronounced central peak and relatively frequent
acceleration bursts up to 80 standard deviations. We note that,
theoretically, time derivative of a dynamical variable does not
necessarily follow the same statistical distribution as that of
the variable.}

Our consideration is restricted to a stationary one-point
distribution function. Two-point statistical analysis is of much
interested and can be made elsewhere.

The paper is organized as follows. In Sec.~\ref{Sec:LDN} we give a
brief description of the 1D LDN model and present the resulting
acceleration distribution, which we treat as a conditional one {by
assigning stochastic properties to certain parameters}. In
Sec.~\ref{Sec:Conditional} we briefly review results of our
previous work and make sample fits of the obtained conditional and
unconditional acceleration distributions {and moments} to the
experimental data. In Sec.~\ref{Sec:Discussion} we discuss the
obtained results and make conclusions.

\section{1D Laval-Dubrulle-Nazarenko model of small-scale turbulence}
\label{Sec:LDN}

In this Section, we present only a brief sketch of the 1D LDN
model and refer the reader to
Refs.~\cite{Laval0101036,Aringazin0305186} for more details; see
also Ref.~\cite{Dubrulle0304035}. This toy model can also be
viewed as a passive scalar in a compressible 1D flow.

{The main assumption of the LDN approach to the 3D Navier-Stokes
turbulence is to introduce and separate large-scale and
small-scale parts in the 3D Navier-Stokes equation by using the
Gabor transformation~\cite{Laval0101036}. This allows to consider
analytically small-scale turbulence coupled to large-scale terms
(the inter-scale coupling). The approach allows one to account for
nonlocal interactions which were argued to be important in
understanding intermittency in developed turbulent flows. The
other, large-scale, part of the equation can be treated separately
(and, in principle, solved numerically given the forcing and
boundary conditions) since the forcing is characterized by
presumably narrow range of small wave numbers, and the small
scales make little effect on it. Small-scale interactions are
modeled by a turbulent viscosity {and were shown numerically to
make small contribution to the anomalous scaling (intermittency)
in the decaying turbulence. Nevertheless, these are important when
fitting model distribution to the experimental data.} The 3D LDN
model of small scale turbulence was used to formulate simplified
1D LDN model, which was studied both in the Eulerian and
Lagrangian frames~\cite{Laval0101036}.}

We use probability density function obtained as a stationary
solution of the Fokker-Planck equation that corresponds to a
consideration of statistically stationary state; statistical
homogeneity and isotropy of the 3D flow is assumed as well. This
equation is derived from the Langevin equation for a component of
Lagrangian acceleration
$a(t)$~\cite{Laval0101036,Aringazin0305186},
\be\label{LangevinLaval}
\frac{\partial a}{\partial t} = (\xi - \nu_{\mathrm t}k^2)a +
\sigma_\perp,
\ee
where $\nu_{\mathrm t} = \sqrt{\nu_0^2+ B^2a^2/k^2}$ is the
turbulent viscosity modeling small-scale interactions, {$\nu_0$ is
constant kinematic viscosity, $B$ is free parameter measuring the
contribution of nonlinearity in $a$ to the turbulent viscosity,
and $k$ is wave number; $\partial_tk = -k \xi$, $k(0)=k_0$, to
model the RDT stretching effect in one-dimensional case.}

In the original 3D LDN model based on the Navier-Stokes equation,
$\xi(t)$ is related to the velocity derivative tensor and
$\sigma_\perp(t)$ describes a forcing of small scales by large
scales via the energy cascade mechanism (nonlocal inter-scale
coupling). In the 1D LDN model, these are approximated by a
sufficiently simple statistics inspired by the Kraichnan ensemble
used for turbulent passive scalar and the Kraichnan-Kazantsev
model of turbulent dynamo: external Gaussian white-in-time noises
along a fluid particle trajectory,
\begin{eqnarray}\label{noises}
\langle\xi(t)\rangle=0, \ \langle\xi(t)\xi(t')\rangle =
2D\delta(t-t'), \ \langle\sigma_\perp(t)\rangle = 0,\ \\ \nonumber
\langle\sigma_\perp(t)\sigma_\perp(t')\rangle =
2\alpha\delta(t-t'), \ \langle\xi(t)\sigma_\perp(t')\rangle =
2\lambda\delta(t-t').
\end{eqnarray}
Here, $D$, $\alpha$, and $\lambda$ are {free parameters measuring
intensity of the noises and their cross-correlation, respectively.
Bigger $D$ and $\alpha$ means bigger contribution of the velocity
derivative tensor and the inter-scale coupling (both viewed here
as short-time autocorrelated processes) to the small-scale
dynamics.}

{The model (\ref{noises})} puts an obvious limitation but is
partially justified by DNS in the laboratory frame of
reference~\cite{Laval0101036}. The averaging is made over ensemble
realizations. Zero means correspond to isotropy of the forces.
{Physically, the small scales are thus assumed to be
stochastically distorted by much larger scales.} We stress that a
correlation between the noises $\xi$ and $\sigma_\perp$ is not ad
hoc assumption but a consequence of their structure as they
contain the same large-scale velocity serving as a unifying agent
between the noises.

It should be emphasized that the 1D LDN toy model and its
particular case (\ref{LangevinLaval}) have several limitations
related to the LDN separation of small and large scales allowing
to study exclusively nonlocal effects associated to the linear
process of distortions of small scales by a strain produced by
large scales, the use of model turbulent viscosity, and
one-dimensionality.

In the Lagrangian frame the wave number $k$ is replaced in terms
of the initial value $k_0=k(0)$ and time while the parameters
acquire dependence on $k_0$~\cite{Laval0101036}; we drop the
subscript 0 in $k_0$ in Eq.~(\ref{LangevinLaval}) and subsequent
formulas to simplify notation.

Thus, one makes a closure by treating the combined effect of large
scales, for which one has a different dynamical LDN equation that
could be in principle solved numerically~\cite{Dubrulle0304035},
and nonlocal inter-scale coupling, as a pair of given external
noises. The large-scale dynamics is local in wave number space and
hence it is weakly affected by the small scales. The price of the
simplification (\ref{noises}) is that one introduces free
parameters to the description. Matching small-scale dynamics to
large-scale one deserves a separate study. Despite 3D turbulence
is known to be more sensitive to large-scale forcing or boundary
conditions, as compared to 2D one, the used simplification
(\ref{noises}) is relevant for high-Reynolds-number flows to some
extent~\cite{Laval0101036,Dubrulle0304035}, and allows one to
advance in analytical treatment of the problem. {It should be
noted that scaling properties of the system described by
Eq.~(\ref{PLaval}) reveal a robust character with respect to the
selection of noises $\xi$ and $\sigma_\perp$ (see
Ref.~\cite{Aringazin0305186} and references therein).}

The acceleration PDF stemming from the stochastic model
(\ref{LangevinLaval})-(\ref{noises}) has been calculated exactly
{in our previous work}~\cite{Aringazin0305186},
\begin{eqnarray}
\label{PLaval}
P(a)
 =C\exp\left[\int_{0}^{a} \!\!\! dx
 \frac{-k^2x\sqrt{\nu_0^2+ B^2x^2/k^2} \!-\! Dx \!+\! \lambda}{Dx^2
-2\lambda x +\alpha}\right]\nonumber\\
 = \frac{C \exp[-{\nu_{\mathrm
t}k^2}/{D}+F(c)+F(-c)]} {(Da^2\!-\!2\lambda a
\!+\!\alpha)^{1/2}(2Bka+\nu_{\mathrm t}k^2)^{{2B\lambda
k}/{D^2}}},
\end{eqnarray}
for constant parameters. Here, $C$ is normalization constant and
we have denoted
\begin{eqnarray}
\label{Fc}
F(c)
 = \frac{c_1k^2}{2c_2D^2c}\ln[\frac{2D^3}{c_1c_2(c-Da+\lambda)}
   \nonumber \\
   \times(
   B^2(\lambda^2 + c\lambda-D\alpha)a
   + c(D\nu_{\mathrm t}^2k^2+c_2\nu_{\mathrm t})
    )
   ],\\
c=-i\sqrt{D\alpha-\lambda^2},\\
c_1 = B^2(4\lambda^3\!+\!4c\lambda^2\! -\!
3D\alpha\lambda-cD\alpha)
    \!+\! D^2(c\!+\!\lambda)\nu_0^2k^2,\\
c_2 = \sqrt{B^2(2\lambda^2 + 2c\lambda-D\alpha)k^2 +
D^2\nu_0^2k^4}.
\end{eqnarray}
The distribution (\ref{PLaval}) is characterized by the presence
of exponential cut off, complicated power-law dependence, and
terms responsible for a skewness (asymmetry with respect to
$a\to-a$).

One way when comparing the model with the experiment is to make a
direct fit of the obtained PDF (\ref{PLaval}) to the experimental
data on unconditional acceleration distribution by assuming all
the parameters and wave number to be constant.

Particularly, this implies a reduction of the original 1D LDN
model since wave number is taken to be fixed so that the
artificial 1D compressibility aimed to model RDT stretching effect
in 1D case is not considered. We note that the Lagrangian
acceleration is usually associated to the dissipative scale, and
in the present paper we do not study dependence of the parameters
on the wave number. Such a dependence for velocity increments was
analyzed in Ref.~\cite{Laval0101036} with the expected result that
for larger scales the velocity increment PDF tends to a Gaussian
form. The Gaussian form is reproduced also when $D \to 0$ and $B
\to 0$, i.e., the process becomes purely additive with a linear
drift term.

Without loss of generality one can put, in a numerical study,
$k=1$ and the additive noise intensity $\alpha=1$ by rescaling the
multiplicative noise intensity $D>0$, the turbulent viscosity
parameter $B>0$, the kinematic viscosity $\nu_0>0$, and the cross
correlation parameter $\lambda$. The particular cases $B=0$ and
$\nu_0=0$ at $\lambda=0$, and the general case at $\lambda=0$ were
studied in detail in Ref.~\cite{Aringazin0305186}. Nonzero
$\lambda$ is responsible for an asymmetry of the PDF
(\ref{PLaval}) and in 3D picture corresponds to a correlation
between stretching and vorticity (the energy cascade).
Particularly, in the Eulerian framework the third-order moment of
spatial velocity increment $\langle (\delta_l u)^3\rangle$ was
found to be proportional to {cross-correlation parameter}, in
accord to a kind of generalized K\'arm\'an-Howarth
relationship~\cite{Laval0101036}.

However, the approximation based on constant parameters does not
allow one to consider both the conditional and unconditional
acceleration statistics.

In the next Section, we extend the model
(\ref{LangevinLaval})-(\ref{PLaval}) by assuming certain model
parameters in Eq.~(\ref{PLaval}) to be dependent on random
velocity fluctuations. This extension is compatible with the 3D
LDN approach as $\xi$ and $\sigma_\perp$ depend on velocity
fluctuations and contain large-scale quantities due to their
definitions~\cite{Laval0101036}. Such a functional dependence and
longtime fluctuations have been ignored when making the
simplification (\ref{noises}). We partially restore them. {This is
the main point of our consideration, and the functional form of
the distribution is thus due to Eq.~(\ref{PLaval}) with certain
parameters being now treated as functions of stochastic velocity
$u$.} Observations are that the acceleration variance does depend
on the same component of velocity fluctuations. Local homogeneity
assumed by Kolmogorov 1941 theory is thus broken that is a
prerequisite to describe turbulence intermittency.  The scaling
approach
indicates an essential character of such a dependence. {Lagrangian
intermittency is known to be much stronger than the Eulerian one
due to existence of very intense vortical structures at small
scales and absence of the so called sweeping effect in the
Lagrangian frame.}

We point out that characteristic time of variation of the
parameters should be sufficiently large to justify approximation
that the resulting PDF (\ref{PLaval}) is used with independent
randomized parameters, $P(a|\textrm{Parameters})$. Two well
separated timescales in the Lagrangian velocity increment
autocorrelation have been established both by experiments and
DNS~\cite{Mordant0206013}. The large timescale has been found of
the order of the Lagrangian integral scale and corresponds to a
magnitude part that is in accord to our assumption that the
intensity of noise along the trajectory is longtime fluctuating.

\section{The conditional acceleration statistics}
\label{Sec:Conditional}

\begin{figure}[tbp!]
\begin{center}
\includegraphics[width=0.45\textwidth]{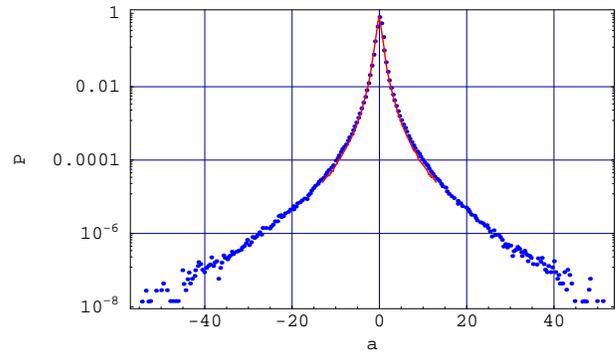}
\caption{\label{Fig1} A comparison of  the experimental
unconditional Lagrangian acceleration PDF (dots) and the
experimental conditional Lagrangian acceleration PDF at velocity
fluctuations $u=0$ (line)~\cite{Bodenschatz2,Mordant0303003}; the
acceleration component $a$ is normalized to unit variance.}
\end{center}
\end{figure}

The experimental unconditional and conditional distributions,
which we denote for brevity by $P_{\mathrm{exp}}(a)$ and
$P_{\mathrm{exp}}(a|u)$ respectively, were found to be
approximately of the same stretched exponential form at $u=0$
(Fig.~\ref{Fig1}), and both reveal a strong Lagrangian turbulence
intermittency~\cite{Mordant0303003}. This similarity indicates
that they share the same process underlying the intermittency.

Accordingly, in our previous
studies~\cite{Aringazin0305186,Aringazin0305459,Aringazin0306022,Aringazin0311098}
we used the result of our direct fit of the PDF (\ref{PLaval}) to
$P_{\mathrm{exp}}(a)$, which was measured with a high precision;
3\% relative uncertainty for $|a|/\langle a^2\rangle^{1/2}\leq
10$~\cite{Bodenschatz2,Mordant0303003}. We assumed that the
parameters $\alpha$ and $\lambda$ entering Eq.~(\ref{PLaval})
depend on the amplitude of Lagrangian velocity fluctuations $u$,
while $D$, $B$, and $\nu_0$ are taken to be fixed at the fitted
values ($k=1$). {Theoretically, only $\alpha$ and $\lambda$ depend
explicitly on large-scale velocity due to 3D LDN model, while the
other parameters not.}

An exponential form of $\alpha(u)$ has been proposed in
Ref.~\cite{Aringazin0305186} and was found to be relevant from
both the (Kolmogorov 1962) phenomenological and experimental
points of view. Particularly, such a form leads to the log-normal
RIN model when $u$ is independent Gaussian distributed with zero
mean~\cite{Aringazin0301245}, and yields the acceleration PDF
whose low-probability tails are in agreement with
experiments~\cite{Beck4-Beck,Aringazin0305186}. Also, we used an
exponential form of $\lambda(u)$ so that the conditional
acceleration PDF (\ref{PLaval}) takes the form
$P(a|u)=P(a|\alpha(u),\lambda(u))$. Such a form was found to
provide good fits of (i) the {conditional probability density
function} $P(a|u)$ to $P_{\mathrm{exp}}(a|u)$; (ii) the
{conditional acceleration variance} $\langle a^2|u\rangle$; and
(iii) the {conditional mean acceleration} $\langle
a|u\rangle$~\cite{Aringazin0306022} at various $u$ that meet the
experimental data~\cite{Mordant0303003}. A brief report on these
results is presented in Ref.~\cite{Aringazin0311098}.

\begin{figure}[tbp!]
\begin{center}
\includegraphics[width=0.45\textwidth]{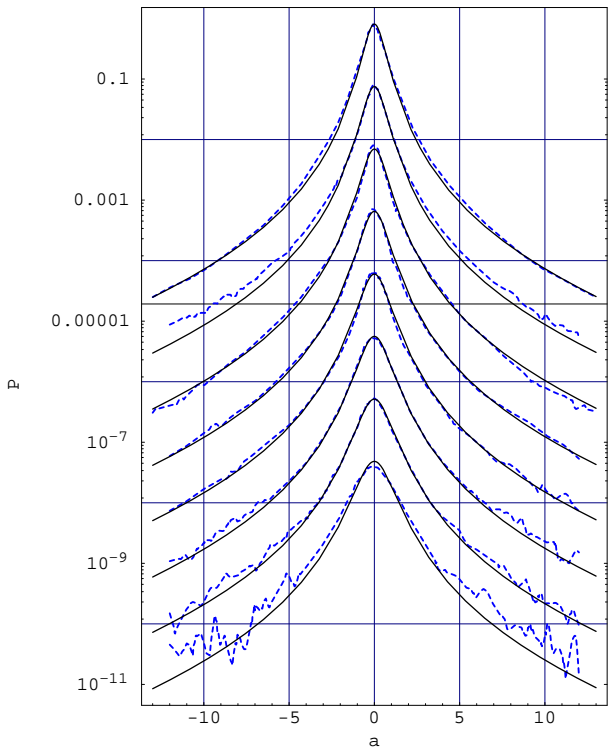}
\caption{\label{Fig2} Theoretical conditional acceleration PDF
$P(a|u)$ (line) and the experimental conditional acceleration PDF
(dashed line) at velocity fluctuations $u=$ 0, 0.45, 0.89, 1.3,
1.8, 2.2, 2.7, 3.1~\cite{Mordant0303003} (from top to bottom,
shifted by repeated factor 0.1 for clarity); the acceleration
component $a$ is normalized to unit variance, and the same
component of velocity $u$ is given in root mean square units.}
\end{center}
\end{figure}

However, a self-consistent consideration of the model assumes
fitting of $P(a|u)$ to $P_{\mathrm{exp}}(a|u)$, and the marginal
PDF computed due to
\be \label{Pmarginal}
P_{\mathrm{m}}(a)=\int_{-\infty}^{\infty}P(a|u)g(u)du,
\ee
where $g(u)$ is PDF of independent velocity fluctuations, should
reproduce $P_{\mathrm{exp}}(a)$. The marginal distribution
corresponds to a convolution of the stationary acceleration
statistics with independent random velocity fluctuations.

In the present paper, we fill this gap. Our task is to fit a
variety of the experimental data, {both on the conditional and
unconditional statistics of acceleration,} with a single set of
fit parameters. For this purpose we use the following natural
steps.

First we fit $P(a|u)=P(a|\alpha(u),\lambda(u))$ given by
Eq.~(\ref{PLaval}) to
$P_{\mathrm{exp}}(a|u)$~\cite{Mordant0303003} assuming that the
parameters depend on $u$ in an exponential way,
\be\label{alphalambda}
\alpha(u) = \alpha_0\exp[{|u|/u_\alpha}],\
\lambda(u)=\lambda_0\exp[{|u|/u_\lambda}].
\ee
Hereafter, we use normalized acceleration $a$ and velocity
fluctuations $u$. The fit parameter set is $D>0$, $\nu_0>0$,
$B>0$, $\lambda_0$, $u_\alpha>0$, and $u_\lambda>0$ ($\alpha_0=1$,
$k=1$). The relations in Eq.~(\ref{alphalambda}) mean that the
additive noise intensity and the correlation between the noises
become higher for bigger velocity fluctuations $|u|$.

We fit $P(a|0)$ to $P_{\mathrm{exp}}(a|0)$, that excludes
$u_\alpha$ and $u_\lambda$ from consideration, by varying $D$,
$\nu_0$, and $B$ at $\alpha_0=1$ and $\lambda_0=-0.005$. We notice
that the available conditional statistics $P_{\mathrm{exp}}(a|u)$
is low for high velocities, the presented acceleration range is
small, $-14 <a<14$, so that a rather big uncertainty remains when
determining fit values of the parameters. Changes in shape of
$P_{\mathrm{exp}}(a|u)$ with $u$ increasing from $u=0$ to $u=3.1$
are captured independently by the fit parameters $u_\alpha$ and
$u_\lambda$. The result is shown in Fig.~\ref{Fig2}. {Good
overlapping of each curve with data points at all fixed magnitudes
of $u$ has been achieved.}

\begin{figure}[tbp!]
\begin{center}
\includegraphics[width=0.45\textwidth]{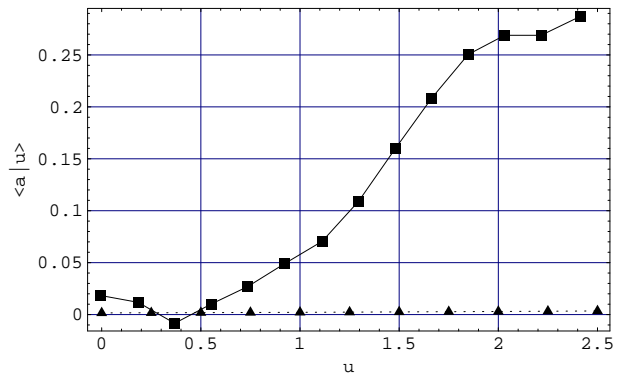}
\caption{\label{Fig3} Theoretical conditional acceleration mean
$\langle a|u\rangle$ (triangles) and the experimental conditional
acceleration mean (squares) as functions of velocity
fluctuations.}
\end{center}
\end{figure}

\begin{figure}[tbp!]
\begin{center}
\includegraphics[width=0.45\textwidth]{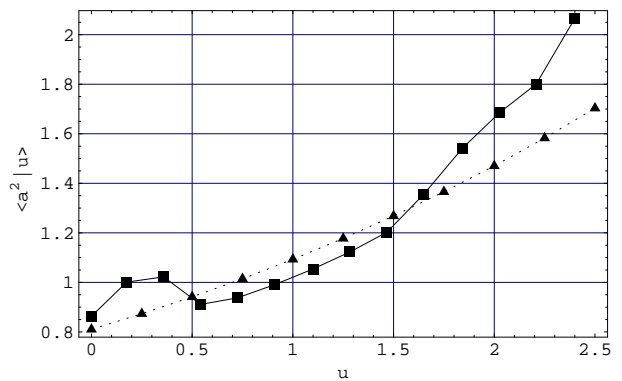}
\caption{\label{Fig4} Theoretical conditional acceleration
variance $\langle a^2|u\rangle$ (triangles) and the experimental
conditional acceleration variance (squares) as functions of
velocity fluctuations.}
\end{center}
\end{figure}

Second we calculate the conditional mean $\langle a|u\rangle$ and
the conditional variance $\langle a^2|u\rangle$ and compare them
with the experimental data. This decreases uncertainty in fit
parameter values. The results are shown in Figs.~\ref{Fig3} and
\ref{Fig4}. Note that $\langle a|u\rangle$ as a function of $u$ is
very small that does not match the experiment. We will discuss
this in Sec.~\ref{Sec:Discussion} below.

\begin{figure}[tbp!]
\begin{center}
\includegraphics[width=0.45\textwidth]{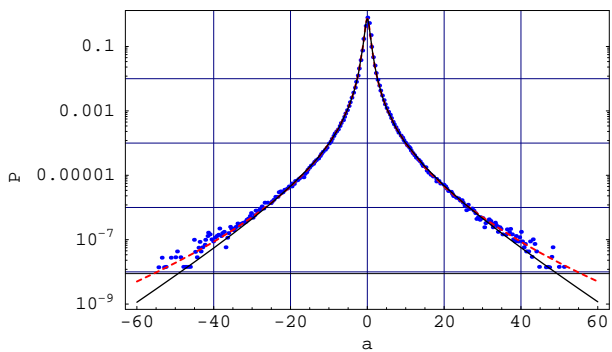}
\caption{\label{Fig5} Theoretical marginal PDF (\ref{Pmarginal})
for Gaussian distributed velocities (line), experimental data at
$R_\lambda=690$~\cite{Bodenschatz2} (dots), and the stretched
exponential fit (\ref{Pexper}) (dashed line).}
\end{center}
\end{figure}

\begin{figure}[tbp!]
\begin{center}
\includegraphics[width=0.45\textwidth]{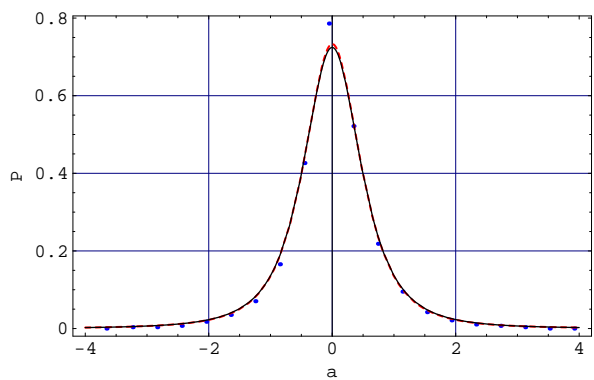}
\caption{\label{Fig6} Lin-lin plot of the central part of the
curves of Fig.~\ref{Fig5}. Same notation as in Fig.~\ref{Fig5}.}
\end{center}
\end{figure}

\begin{figure}[tbp!]
\begin{center}
\includegraphics[width=0.45\textwidth]{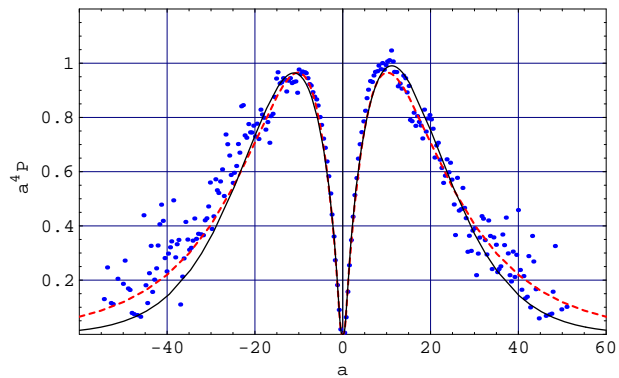}
\caption{\label{Fig7} Contribution to the kurtosis, $a^4P(a)$.
Same notation as in Fig.~\ref{Fig5}.}
\end{center}
\end{figure}

Finally we calculate numerically the marginal distribution
$P_{\mathrm{m}}(a)$ given by Eq.~(\ref{Pmarginal}) with the
conditional PDF $P(a|\alpha(u),\lambda(u))$ and Gaussian
distribution of velocity fluctuations,
\be\label{Gaussianu}
g(u)=\frac{1}{\sqrt{2\pi}}\exp\left[-\frac{u^2}{2}\right],
\ee
at fixed $a$ ranging from $-100$ to 100 with the step 0.1. Then we
make an interpolation and fit it to $P_{\mathrm{exp}}(a)$. A
noticeable effect of the integration over $u$ with Gaussian $g(u)$
is a widening of tails of the distribution that meets
Fig.~\ref{Fig1}; the integration range $-20\leq u\leq 20$ has been
used. The fit of $P_{\mathrm{m}}(a)$ to $P_{\mathrm{exp}}(a)$
strongly decreases the uncertainty but the most strict
determination of fit values comes due to a comparison of the
theoretical contribution to fourth-order moment, $a^4P(a)$, with
the experimental data. The results are shown in Figs.~\ref{Fig5},
\ref{Fig6}, and \ref{Fig7}. Quality of these sample fits is better
than in the other recent stochastic models reviewed in
Ref.~\cite{Aringazin0305186}. In particular, the core of the
unconditional distribution reproduces very well that given by the
stretched exponential (\ref{Pexper}) as shown in Fig.~\ref{Fig6}.
However, both curves a bit underestimate the height at $a=0$.

The value $\lambda_0=-0.005$ has been obtained by adjusting the
theoretical curve to slightly different heights of the peaks of
the observed $a^4P(a)$ shown in Fig.~\ref{Fig7}. Note that the
model does not assume the use of ad hoc skewness of the forcing.
Nonzero cross correlation parameter $\lambda$ naturally results
not only in small mean acceleration but also in a skewness of both
the theoretical distributions $P(a|u)$ and $P_{\mathrm{m}}(a)$.
This skewness may be associated to the Eulerian downscale skewness
generation, which despite of being small for homogeneous flows is
known to be of a fundamental character in the inertial range
(Kolmogorov four-fifths law), since the Eulerian $\langle(\delta_l
u)^3\rangle$ was found to be proportional to cross-correlation
parameter.

{We stress that the observed very small skewness of acceleration
distribution is attributed to the effect of anisotropy of the
studied flow. How the large-scale asymmetry affects smallest
scales of the flow is an interesting problem. Our fit made by
using nonzero $\lambda$ is of an illustrative character, to verify
whether it can explain the observed increase of the conditional
mean acceleration with increasing velocity {depicted in
Fig.~\ref{Fig3}}. This issue will be discussed further in
Sec.~\ref{Sec:Discussion}.}

The following remarks are in order. Our finding is that the
condition $u_\alpha \leq u_\lambda$ provides a convergence of
$P_{\mathrm{m}}(a)$. Also, $u_\lambda$ should not be small to
provide assumed condition $\lambda \ll \alpha$ at arbitrary $u$
(the cross correlation is small as compared to both noise
intensities $\alpha$ and
$D$)~\cite{Aringazin0305186,Laval0101036}. We used these criteria
when making the fits.

The resulting sample fit values are given by
\begin{eqnarray}\label{fits}
D=2.1, \ \nu_0=5.0, \ B=0.35,\\
\nonumber \lambda_0=-0.005, \ u_\alpha=3.0,\ u_\lambda=3.0,
\end{eqnarray}
with $\alpha_0=1$ and $k=1$. The theoretical curves in
Figs.~\ref{Fig2}--\ref{Fig7} are shown for this sample set of
values, which require a further fine tuning. Such a small value of
$\lambda$ as compared to $\alpha$ or $D$ is in agreement with that
obtained in the LDN direct numerical simulations. The calculated
flatness factor $F=49.3$ of $P_{\mathrm{m}}(a)$ is in agreement
with the experimental value 55 $\pm$ 8.

To summarize, the considered Navier-Stokes equation based 1D toy
model (\ref{PLaval})-(\ref{alphalambda}) is capable to fit all the
available high-precision experimental data on the conditional and
unconditional Lagrangian acceleration
statistics~\cite{Bodenschatz,Bodenschatz2,Mordant0303003} with the
single set of parameters (\ref{fits}) to a good accuracy, with an
exception being only the conditional mean acceleration.

\section{Discussion and conclusions}
\label{Sec:Discussion}

One can see from Fig.~\ref{Fig3} that at the values of fit
parameters (\ref{fits}) the predicted conditional mean
acceleration $\langle a|u\rangle$ qualitatively is in agreement
but does not reproduce the experimental data. Namely, it is
nonzero due to nonzero $\lambda$ and increases with the increase
of $|u|$ but remains to be very small even at high values of
$|u|$. The conditional mean acceleration is evidently zero for a
symmetrical distribution ($\lambda=0$) and should be zero for
statistically homogeneous isotropic turbulence. The observed
departure from zero is thought to reflect anisotropy of the
studied flow albeit the DNS of homogeneous isotropic turbulence
also reveals slightly nonzero mean~\cite{Mordant0303003}.

To reduce the discrepancy, we have tried the value $u_\lambda=1.0$
instead of $u_\lambda=3.0$ to provide faster increase of
$|\lambda|$ for higher $|u|$. This implies a good fit to the
experimental conditional mean acceleration (see, e.g., Fig.~2 in
Ref.~\cite{Aringazin0311098}) but we found an excess asymmetry of
$P(a|u)$ at high $u$, with big departure from observations, and
divergencies when calculating $P_{\mathrm{m}}(a)$. The reason of
the divergency is that $\lambda(u)$ at $u_\lambda=1.0$ grows
faster than $\alpha(u)$ at $u_\alpha=3.0$ so that $\lambda$
becomes comparable or bigger than $\alpha$ with increasing $u$,
and when $\lambda^2 \to D\alpha$ the function $F(c)$ defined by
Eq.~(\ref{Fc}) undergoes unbound growth. Thus we conclude that the
observed conditional mean acceleration is mainly due to the flow
anisotropy effect rather than some intrinsic dynamical mechanism
associated to the developed turbulence.

In general one observes a rather small relative increase of the
conditional mean acceleration for higher $|u|$ that eventually
reflects a coupling of the acceleration to large scales of the
studied flow~\cite{Laval0101036,Chevillard0310105}. This coupling
could be accounted also by introducing a correlation between the
acceleration and velocity fluctuations. This possibility is of
much interest to explore as it may yield the deficient increase of
$\langle a|u\rangle$ but it is beyond the scope of the present
formalism, which assumes an independent velocity statistics. We
note also that in contrast to the experimental data on the
variance $\langle a^2|u\rangle$ the experimental $\langle
a|u\rangle$ exhibits small asymmetry with respect to $u\to -u$
(not shown in Fig.~\ref{Fig3}).

In the present paper, the multiplicative noise intensity $D$ was
taken to be independent on the velocity fluctuations $u$. The
effect of variation of $D$ has been considered in
Ref.~\cite{Aringazin0305186} with the qualitative result that it
does not provide the specific change in shape of $P(a|u)$ observed
in experiments. However, a weak dependence of $D$ on $u$ can not
be ruled out.

In summary, the presented 1D LDN type stochastic toy model with
the velocity-dependent additive noise intensity and cross
correlation parameter is shown to capture main features of the
observed conditional and unconditional Lagrangian acceleration
statistics to a good accuracy except for the discrepancy in the
conditional mean acceleration which can be attributed to certain
coupling of the acceleration to large scales of the studied flow.

The main result is of course not only good sample fits which are
important to test performance of the model but also certain
advance in understanding of the mechanism of Lagrangian
intermittency provided by the dynamical Laval-Dubrulle-Nazarenko
approach to small-scale turbulence.

The central point is that the LDN toy model has a strong deductive
support from the Navier-Stokes turbulence. The obtained exact
analytic result for the conditional acceleration distribution and
the use of recent high-precision Lagrangian experimental data on
conditional and unconditional acceleration statistics provide a
detailed analysis of the mechanism within the adopted framework.
Effects of large timescales (nonlocality) and turbulent viscosity
(nonlinearity) have been found of much importance in Lagrangian
acceleration steady-state statistics. The detailed study of
conditional acceleration statistics have revealed a specific model
structure of the external large-scale dynamics and nonlocal
inter-scale coupling for homogeneous high-Reynolds-number flows.
The additive noise associated to the downscale energy transfer
mechanism encodes the main contribution to the velocity dependence
of the acceleration statistics. The cross correlation between the
model additive and multiplicative noises associated to a
correlation between stretching and vorticity naturally provides a
skewness of distributions and a nonzero mean. Weakness of this
correlation is a theoretical requirement that meets the Lagrangian
and Eulerian experiments and DNS of homogeneous isotropic
turbulence. The observed conditional mean acceleration is mainly
related to the flow anisotropy. The cross correlation is related
to the four-fifths Kolmogorov law but the effect of skewness is
negligibly small as the result of relatively large intensity of
the additive noise, which tends to symmetrize acceleration
distributions. This is a dynamical evidence implied by the model
rather than a direct consequence of a priori assumption on
isotropy in the spirit of Kolmogorov 1941 theory. The use of
exponential dependence of certain noise parameters on
statistically independent Gaussian distributed Lagrangian velocity
fluctuations has been found appropriate to cover new experimental
data on conditional statistics and to transfer from the
conditional to unconditional acceleration distribution both
exhibiting a strong Lagrangian intermittency of the flow. Such a
dependence is also compatible with the log-normal statistics
assumed by the Kolmogorov 1962 theory. The Gaussian white-in-time
multiplicative noise and longtime correlated intensity of the
additive noise were both found to make an essential contribution
to intermittent bursts.

\acknowledgments{The author is grateful to A.M. Reynolds for
valuable comments on issues related to the present formalism and
sending his work, and M.I. Mazhitov for stimulating discussions.}


\end{document}